\documentclass[submission]{eptcs}
\providecommand{\event}{TARK 2017} 

 \usepackage{amsmath,amssymb,graphicx}
 \newcommand{\citeyear}{\cite}
\newcommand{\prf}{\noindent{\bf Proof:} }
\newcommand{\eprf}{\bbox\vspace{0.1in}}
\newcommand{\bbox}{\vrule height7pt width4pt depth1pt}
\newtheorem{theorem}{Theorem}[section]

\newtheorem{proposition}[theorem]{Proposition}
\newtheorem{definition}[theorem]{Definition}

\newcommand{\IR}{\mbox{$I\!\!R$}}

\newcommand{\BEQ}{\begin{equation}}
\newcommand{\EEQ}{\end{equation}}
\newcommand{\BEA}{\begin{eqnarray}}
\newcommand{\EEA}{\end{eqnarray}}
\newcommand{\BA}{\begin{align}}
\newcommand{\EA}{\end{align}}
\newcommand{\BIT}{\begin{itemize}}
\newcommand{\EIT}{\end{itemize}}
\newcommand{\BNUM}{\begin{enumerate}}
\newcommand{\ENUM}{\end{enumerate}}

\newcommand{\fullv}[1]{#1}
\newcommand{\shortv}[1]{}
\newcommand{\posted}[1]{}

\title{Games With Tolerant Players}
\author{Arpita Ghosh
\institute{Cornell University\\
Ithaca, NY 14853, USA}
\email{ag865@cornell.edu}
\and
Joseph Y. Halpern
\institute{Cornell University\\
Ithaca, NY 14853, USA}
\email{halpern@cs.cornell.edu}
}
\def\titlerunning{Games With Tolerant Players}
\def\authorrunning{A. Ghosh \& J. Y. Halpern}

\begin{document}

\maketitle

\begin{abstract}
A notion of \emph{$\pi$-tolerant equilibrium} is defined that takes
into account that players have some tolerance regarding payoffs in a
game.  This solution concept generalizes Nash
and refines $\epsilon$-Nash equilibrium in a natural way.  We show that
$\pi$-tolerant equilibrium can explain cooperation in social dilemmas such as
Prisoner's Dilemma and the Public Good game.  We then examine
the structure of \emph{particularly cooperative $\pi$-tolerant equilibria},
where players are as cooperative as they can be, subject to their
tolerances, in Prisoner's Dilemma.   To the extent that cooperation is
due to tolerance, these results provide guidance to a mechanism designer
who has some control over the payoffs in a game, and suggest ways in which
cooperation can be increased.
\end{abstract}


\section{Introduction}
People exhibit systematic deviations from the
predictions of game theory.  For example, they do not always act so
as to maximize their expected utility in games such as Prisoner's
Dilemma (to the extent that their utility is accurately  characterized
by the payoffs in the game).  Many alternative models have been
proposed to explain these deviations; the explanations range from
players having \emph{other-regarding} preferences, so that 
they prefer to avoid inequity, or prefer to maximize social welfare
(see, e.g.,  \cite{Bo-Oc,Ch-Ra,Fe-Sc})
to \emph{quantal-response equilibrium}, which assumes that,
with some probability, players make mistakes and do not play
rationally \cite{MK-Pa95}.

The literature here is enormous; a complete bibliography would
be almost as long as this paper.  Nevertheless, we propose yet another
approach to explaining what people do.  Our explanation assumes that
people have some \emph{tolerance} for not getting the optimal payoff;
the degree of tolerance is measured by how far from an optimal payoff
they find acceptable.  This idea is certainly not new: it is implicit
in notions like $\epsilon$-equilibrium and \emph{satisficing}
\cite{Simon55,Simon56}, although the details are different.
Moreover, it is clear that people do have some tolerance.  There are
many potential reasons for this.  First, although we often identify
payoffs and utilities, the payoffs in a game may not
represent a player's true utility; a player may in fact be indifferent
between receiving a payoff of $a$ and $a-t$ if $t$ is sufficiently
small. (This is in the spirit of the ``satisficing'' view.)
Or there may be a recommended strategy, and some overhead in
switching (which again is not being captured in the game's payoffs).
\posted{
Or players might work at a coarse level, using a language that can't
distinguish situations that a modeler might think should get different
utility (cf.~\cite{BHP13}).} 
For whatever reason, it seems reasonable to assume that players may
have some tolerance regarding payoffs.
However, there is no reason to believe that all players have the same
tolerance.
We thus assume that there is a distribution over possible tolerances for
each player,
captured by a profile $\pi = (\pi_1, \ldots,
\pi_n)$ of distributions, where $\pi_i$ is a distribution over the
possible tolerances of player $i$.

Intuitively, we imagine
that we have a large population of players who could be player $i$;
if we choose player $i$ at random from this poplulation, then  with probability
$\pi_i(t)$, she will have
tolerance $t$.
\fullv{
(Of course, in many applications, it is reasonable to
assume that all players are chosen from the same pool, so that $\pi_1 =
\cdots = \pi_n$.)
We can relate this to more traditional
game-theoretic considerations by thinking of these tolerances as representing
different \emph{types} of player $i$; that is, a type is associated
with a tolerance.  There is some psychological evidence to support
this viewpoint; specifically,  there some evidence that whether
someone uses satisficing-style behavior, and the extent to which it is
used, is a personality trait, with a strong genetic component that endures over
time \cite{SS11}.
}
In this setting, we define an equilibrium notion that we call
\emph{$\pi$-tolerant equilibrium}.  A profile
$\sigma$ of possibly mixed strategies, one
for each player, is a $\pi$-tolerant equilibrium if, roughly speaking,
for each type $t$ of player $i$ (i.e., each possible tolerance that
player $i$ can have), we can assign a
mixed strategy to type $t$ in such a way that (1) each
of the pure strategies in the mixture is a best response to
$\sigma_{-i}$ (i.e., what the other players are doing) and (2)
$\sigma_i$ represents the convex combination of what all the
types of player $i$ are doing.  Intuitively, the other players don't
know what type of player $i$ they are facing; $\sigma_i$ describes the
aggregate behavior over all types of player $i$.  We can show that a
Nash equilibrium is a 0-tolerant equilibrium (i.e., if we take $\pi_1 =
\cdots = \pi_n$ to be the distribution that assigns probability 1 to
players having tolerance 0); moreover, every Nash equilibrium is a
$\pi$-tolerant equilibrium for all $\pi$.  Similarly,
if $\pi_1^\epsilon = \cdots = \pi_n^\epsilon$ all asign probability
1 to players having tolerance $\epsilon$, then a
$\pi^{\epsilon}$-tolerant equilibrium is an $\epsilon$-Nash equilibrium.
(The converse is not quite true; see Section~\ref{sec:toleranteq}.)

After defining $\pi$-tolerant equilibria in
Section~\ref{sec:toleranteq}, in Section~\ref{sec:dilemma},
we review the definition of social
dilemmas, discuss the observed behavior in social dilemmas that
we seek to explain, and show how tolerance
can explain it.

Our interest in social dilemmas is only part of why we are interested
in tolerance.
We are also interested in taking advantage of tolerance when designing
mechanisms.  We illustrate the potential in Section~\ref{sec:PD} by
investigating this issue in the context of
Prisoner's Dilemma.  Although
Prisoner's Dilemma may seem to be a limited domain, it can model a
range of two-player interactions with appropriate meanings
ascribed to the actions of cooperating and defecting.
Our 
analysis of Prisoner's Dilemma with tolerance isolates the
factors that determine the equilibrium level of cooperation in the
game, providing guidelines  (to the extent to which tolerance is
indeed the explanation for observed cooperation) for how a designer,
who may be able to modify or control the payoffs from certain actions,
can adjust them to achieve particular levels of cooperative behavior
in equilibrium.

\section{$\pi$-Tolerant Equilibrium}\label{sec:toleranteq}
We consider normal-form games here.
A  \emph{normal-form game} is a tuple $\Gamma = (N, (S_{i})_{i
  \in N}, (u_i)_{i \in N})$,
where $N$ is a finite set of \emph{players}, which for convenience we
take to be the set $\{1, \ldots, n\}$, $S_{i}$ is the set of
pure strategies available to player $i$, 
which for convenience we also take to be finite,
and $u_i$ is $i$'s utility
function.  As usual, we take
$S =  S_{1} \times \cdots \times S_n$.
A \emph{mixed strategy} for player $i$ is a distribution on $S_i$.
Let $\Sigma_i$ denote the mixed strategies for player $i$, and let
$\Sigma = \Sigma_1 \times  \cdots \times \Sigma_n$.
Elements of $\Sigma$ are called \emph{mixed-strategy
  profiles}; given $\sigma \in \Sigma$, we denote by $\sigma_{i}$ the
$i$th component of the tuple $\sigma$, and by $\sigma_{-i}$ the
element of $\Sigma_{-i}$ consisting of all but the $i$th component of
$\sigma$. The utility function $u_i: S \rightarrow \IR$; that is,
$u_i$ associates with each pure strategy profile a real number, which
we can think of as $i$'s utility.  We can extend $u_i$ to $\Sigma$ in
the obvious way, by linearity.

We take $T_i$ to be the set of possible tolerances for player $i$.
Each element of $T_i$ is a non-negative real number.  For simplicity in this
discussion, we take $T_i$ to be finite, although the definitions that we are
about to give go through with minimal change if $T_i$ is infinite
(typically, summations have to be changed to integrations).
We identify a tolerance with a \emph{type}; it can be viewed as
private information about player $i$.
Let $\pi_i$ be a distribution on $T_i$, the set of possible types of
$i$ (under this viewpoint), and let $\pi = (\pi_1, \ldots,
\pi_n)$. 

We want to define what it means for a mixed-strategy profile
$(\sigma_1, ... \sigma_n)$ to be a $\pi$-tolerant equilibrium.   The
intuition is that $\sigma_i$ represents a population distribution.  If
$\sigma_i$ puts probability $p_i$ on the pure strategy $s_i$, then a
fraction $p_i$ of the population
(of agents who could play the role of player $i$)
plays $s_i$.  Similarly, if
$\pi_i$ puts a probability $p_i$ on a tolerance $t$, then a fraction
$p_i$ of the population of agents who could be player $i$ has type $t$.
Given our view that a mixed strategy for player $i$ really represents
a population of players each playing a pure strategy,
in a $\pi$-tolerant
equilibrium, we want all players of tolerance $t$ to be playing a
mixed strategy such that each strategy in the support is within a
tolerance $t$ of being a best response to what the other players are
doing.%
\footnote{We should stress that although we view a mixed strategy for
  player $i$ as representing a population of players, each playing a
  pure strategy, nothing in the formal definitions requires this.
  There could equally well be a single player $i$ playing a mixed strategy.}

\begin{definition}\label{def:consistent} {\rm
A pure strategy  $s_i$ for player $i$ is \emph{consistent with
a tolerance $t$ for player $i$ and a mixed-strategy profile
$\sigma_{-i}$} for the players other than $i$  if $s_i$ is a $t$-best
response
to $\sigma_{-i}$;
that is, if, for all strategies $s_i'$ for player $i$,
$$u_i(s_i',\sigma_{-i}) \le u_i(s_i,\sigma_{-i}) + t.$$}
\end{definition}

\begin{definition} {\rm $\sigma$ is a
  \emph{$\pi$-tolerant equilibrium} if, for each player $i$, there is
  a mapping $g_i$ from the set $T_i$ of possible tolerances of player
  $i$ to mixed strategies for 
player $i$ such that the following conditions hold:
\begin{itemize}
\item[E1.] The support of $g_i(t)$
consists of only pure strategies that are consistent with $t$ and
$\sigma_{-i}$.
(Intuitively, a player $i$ of type $t$ will play only strategies
that are $t$-best responses to $\sigma_{-i}$.)
\item[E2.] $\sum_t \pi_i(t) g_i(t) = \sigma_i$.%
\end{itemize}
}
\end{definition}
Note that if $(\sigma_1, ... \sigma_n)$ is a $\pi$-tolerant
equilibrium, then there might not be any type of player $i$ that plays
strategy $\sigma_i$.  Rather, $\sigma_i$ describes the other
players' perception of what a ``random'' instance of player $i$ is doing.
Thus, if player $i$ has two possible types, say $t$ and
$t'$, where $t$ occurs with probability $1/3$ and $t'$
occurs with probability $2/3$, then E2 requires that $\sigma_i =
\frac{1}{3}g_i(t) + \frac{2}{3}g_i(t')$.

Every Nash equilibrium is clearly a $\pi$-tolerant
equilibrium for all $\pi$:  For if $\sigma$ is a Nash equilibrium,
then each pure strategy in the support of $\sigma_i$ is a best
response to $\sigma_{-i}$, so must be consistent with $t$ and
$\sigma_{-i}$ for all types $t$.  Thus, if we take $g_i(t) = \sigma_i$
for all $t$, then E1 and E2 above are clearly satisfied.  Moreover,
the Nash equilibria are precisely the $\delta^0$-tolerant equilibria, where
$\delta^0 = (\delta_1^0, \ldots, \delta_n^0)$ and $\delta_i^0$ puts
probability 1 on type 0.

It is similarly easy to check that if
$\delta^\epsilon = (\delta_1^\epsilon, \ldots, \delta_n^\epsilon)$,
where $\delta_i^\epsilon$ puts probability 1 on type $\epsilon$, then
every $\delta^\epsilon$-tolerant equilibrium is an $\epsilon$-Nash
equilibrium.  The converse is \emph{not} true, at least not the way
that $\epsilon$-Nash is typically defined
(see, e.g., \cite{M97}).

For example, consider
Prisoner's Dilemma.  As is well known, defecting is the dominant
strategy.  Given $\epsilon > 0$, there exists a $\delta > 0$
sufficiently small such that the
mixed strategy $\delta C + (1-\delta) D$ (cooperating with probability
$\delta$ and defecting with probability $1-\delta$) is an
$\epsilon$-best response no matter what the other player does; thus,
both players using this strategy is an $\epsilon$-Nash equilibrium.
However, it is not a $\delta^\epsilon$-tolerant equilibrium if $C$
is not an $\epsilon$-best response.
Interestingly, Goldberg and Papadimitriou \citeyear{GP06}
defined a (nonstandard) notion of $\epsilon$-Nash equilibrium
where all
strategies in the support of a mixed strategy are required to be
$\epsilon$-best responses.  This corresponds exactly to our notion of
$\delta^\epsilon$-tolerant equilibrium.

Thus, $\pi$-tolerant equilibrium refines Nash equilibrium and
$\epsilon$-Nash
equilibrium in an arguably natural way that allows for beliefs
regarding agents' tolerance.
We can also view it as a generalization of $\epsilon$-Bayes-Nash equilibrium
in Bayesian games.  Recall that in a Bayesian game, each
  player $i$ has a type in a set $T_i$.  It is typically assumed that
  there is a (commonly known) distribution over $T = T_1 \cdots \times
  \cdots T_n$, and that a player's utility can depend on his type.
The notion of \emph{$\epsilon$-Bayes-Nash equilibrium} in a
Bayesian game a natural extension of $\epsilon$-Nash equilibrium.  If
we take a player's 
type to be his tolerance, and take all types to agree on the
utilities, then a $\pi$-tolerant equilibrium is a $\epsilon$-Bayes-Nash
equilibrium in the sense of the Goldberg-Papadimitriou definition,
\emph{provided that the $\epsilon$ can depend on the player's type}.
That is, rather than having a uniform $\epsilon$, we have a
type-dependent $\epsilon$.   We believe that focusing on
tolerance and its consequences gives more insight than thinking in
terms of this nonstandard notion of Bayes-Nash equilibrium; that is
what we do in the remainder of the paper.


We conclude this section by showing that greater tolerance leads to
more equilibria.
While this is intuitively clear, the proof 
\fullv{(which can be found in the appendix)}
is surprisingly nontrivial.  Given a distribution $\pi_i$, let $F^{\pi_i}$
denote 
the corresponding cumulative distribution; that is, $F^{\pi_i}(t) =
\sum_{t' \le t} \pi_i(t')$.  Say that $\pi_i'$ \emph{stochastically
  dominates} $\pi_i$ if $F^{\pi_i'} \le F^{\pi_i}$; that is,
  $F^{\pi_i'}(t) \le F^{\pi_i}(t)$ for all $t$.  Thus, the probability
of getting greater than $t$ with $\pi_i'$ is at least as high as the
probability of getting greater than $t$ with $\pi_i$.  Intuitively, $\pi_i'$
stochastically dominates $\pi_i$ if $\pi_i'$ is the result of shifting
$\pi_i$ to the right.  A profile $\pi'= (\pi'_1, \ldots,
\pi_n')$ stochastically dominates $\pi = (\pi_1, \ldots, \pi_n)$ if
$\pi_i'$ stochastically dominates $\pi_i$ for all $i$.

\begin{theorem}\label{thm:stochastic} If $\pi'$ stochastically
  dominates $\pi$, then 
every $\pi$-tolerant equilibrium is a $\pi'$-tolerant equilibrium.
\end{theorem}

\prf See the appendix.
\eprf


\section{Social Dilemmas}\label{sec:dilemma}

Social dilemmas are situations in which there is a tension
between the collective interest and individual interests: every
individual has an
incentive to deviate from the common good and act selfishly, but if
everyone deviates, then they are all worse off.
Following Capraro and Halpern \citeyear{CH2014},
we formally define a social dilemma
as a normal-form game with a unique Nash equilibrium $\sigma^N$ and a unique
welfare-maximizing profile $s^W$, both pure strategy profiles,
such that each player's expected utility if $s^W$ is played is higher
than his utility if $s^N$ is played.
While this is
a quite restricted set of games, it includes many
of the best-studied games in the game-theory literature.

We examine the same four games as Capraro and
Halpern~\citeyear{CH2014}, and 
show that \fullv{the} experimentally observed regularities in these games
can \fullv{also} be explained using tolerance.%
\footnote{The description of the 
games and   observations is taken
almost verbatim  from Capraro and Halpern~\citeyear{CH2014}.}

\begin{description}
\item[\textbf{Prisoner's Dilemma.}]  Two players can either cooperate
($C$) or defect ($D$).  To relate our results to experimental results on
Prisoner's Dilemma, we consider a subclass
of Prisoner's Dilemma games where we think of cooperation as meaning
that a player pays a cost $c > 0$ to give a benefit $b>c$ to the other
player.  If a player defects, he pays nothing
and gives nothing.  Thus, the payoff of $(D,D)$ is
$(0,0)$, the payoff of $(C,C)$ is $(b-c,b-c)$, and the payoffs of $(D,C)$ and
$(C,D)$ are $(b,-c)$ and $(-c,b)$, respectively.
The condition $b>c$ implies that $(D,D)$ is the unique Nash equilibrium
and $(C,C)$ is the unique welfare-maximizing profile.

\item[\textbf{Traveler's Dilemma.}] Two travelers have identical
luggage, which is damaged (in an identical way) by an airline.  The
airline offers to recompense them for their luggage.
They may ask for any dollar amount between $L$ and $H$
(where $L$ and $H$ are both positive integers).
There is only one catch.  If they ask for the same amount, then that is
what they will both receive.  However, if they ask for different
amounts---say one asks
for $m$ and the other for $m'$, with $m < m'$---then whoever asks
for $m$ (the lower amount) will get $m+b$ ($m$ and a bonus of $b$),
while the other player gets $m-b$: the lower amount and a penalty of
$b$.  $(L,L)$ is thus
the unique Nash equilibrium, while $(H,H)$ maximizes social welfare,
independent of $b$.

\item[\textbf{Public Goods game.}] $N\geq2$ contributors are endowed
with 1 dollar  each; they must simultaneously decide how much, if
anything, to contribute to a public pool.
(The contributions must be in whole cent amounts.)
The total contribution pot is then multiplied by a
constant strictly between 1 and $N$,
and then evenly redistributed among all players. So
the payoff of player $i$ is
$u_i(x_1,\ldots,x_N)=1-x_i+\rho(x_1+\ldots+x_N)$, where
$x_i$ denotes $i$'s contribution, and $\rho\in\left(\frac1N,1\right)$
is the \emph{marginal return}.
(Thus, the pool is multiplied by $\rho N$ before being split
evenly among all players.)  Everyone contributing nothing to
the pool is the unique Nash equilibrium, and everyone contributing their
whole endowment to the pool is the unique welfare-maximizing profile.

\item[\textbf{Bertrand Competition.}] $N\geq2$ firms compete to sell
their identical product  at a price between the ``price floor'' $L\geq 2$
and the ``reservation value'' $H$.
(Again, we assume that $H$ and $L$ are integers, and all prices must
be integers.)   The firm that chooses the lowest
price, say $s$, sells the product at that price, getting a payoff of
$s$, while all other firms get a payoff of 0. If there are ties,
then the sales are split equally among all firms that choose the lowest
price.   Now everyone choosing $L$ is the unique Nash equilibrium, and
everyone choosing $H$ is the unique welfare-maximizing profile.%
\footnote{We require that $L \ge 2$ for otherwise we would not have a
  unique Nash equilibrium, a condition we imposed on Social Dilemmas.
If $L = 1$ and $N=2$, we get two Nash
  equilibria: $(2,2)$ and $(1,1)$; similarly, for $L=0$, we also get
  multiple Nash equilibria, for all values of $N \ge 2$.}

\end{description}

From here on, we say that a player \emph{cooperates} if he plays
his part of the socially-welfare maximizing strategy profile and
\emph{defects} if he plays his part of the Nash equilibrium strategy profile.
While Nash equilibrium predicts that people should always defect in
social dilemmas, in practice, we see a great deal of cooperative
behavior.   But the cooperative behavior exhibits a great deal of
regularity.  Here are some regularities that have been observed
(although it should be noted that in some cases the evidence is rather
limited---see the discussion of Bertrand Competition at the end of
this section):
\begin{itemize}
\item The degree of cooperation in the Prisoner's dilemma depends
positively on the benefit of mutual cooperation and negatively on the
cost of cooperation \cite{capraro2014heuristics,EZ,Rapoport}.
\item The degree of cooperation in the Traveler's Dilemma depends
negatively on the bonus/penalty \cite{CGGH99}.
\item The degree of cooperation in the Public Goods game depends
positively on the constant marginal return \cite{Gu,IWT}.
\item The degree of cooperation in the Public Goods game depends
positively on the number of players \cite{IWW,Ze}.
\item The degree of cooperation in the Bertrand Competition depends
negatively on the number of players \cite{DG00}.
\item The degree of cooperation in the Bertrand Competition depends
  negatively on the price floor \cite{D07}.
\end{itemize}

Of course, as mentioned in the introduction, there have been many
attempts to explain the regularities that have been observed in social
dilemmas.  However, very few can
actually explain all the regularities mentioned above.  Indeed,
the only approaches seem to be Charness and Rabin's \citeyear{Ch-Ra}
approach, which assumes that agents care about maximizing social
welfare and the utility of the worst-off individual as well as their
own utility, and the translucency approach introduced by Halpern and
Pass \citeyear{HaPa13} and adapted by Capraro and Halpern
\citeyear{CH2014} to explain social dilemmas: roughly
speaking, a player is translucent to the degree that he believes that,
with some probability, other players will know what he is about to do. 

To show how tolerance can explain cooperation in social dilemmas,
we first examine the relationship between tolerance and the parameters
of the various social
dilemmas we are considering.  We consider two settings.  In the first,
we ask when it is consistent for a player $i$ of type $t$ to cooperate. In
the second, we ask when it is rational for a player $i$ of type $t$ who 
believes (as assumed by Capraro and Halpern \citeyear{CH2014}) that
each other player $j$ is playing $\beta s_j^W + (1-\beta)s_j^N$ to cooperate
(i.e., $i$ believes that each other player is cooperating with
probability $\beta$, 
defecting with probability $(1-\beta)$, and not putting positive
probability on any other strategy).  We write $\beta s_{-i}^W +
(1-\beta) s_{-i}^N$ for the corresponding mixed-strategy profile.
Say that a player has type
$(t,\beta)$ in the second case; in the spirit of
Definition~\ref{def:consistent}, say that cooperation is
\emph{consistent  with type $(t,\beta)$ for player $i$} if
for all strategies $s_i'$ for player $i$,
$$u_i(s_i',\beta s_{-i}^W +
(1-\beta) s_{-i}^N) \le u_i(s_i^W,\beta s_{-i}^W +
(1-\beta) s_{-i}^N) + t.$$

For a Prisoner's
Dilemma of the form described in Section~\ref{sec:dilemma},
consistency is independent of the strategy the other player is using
(and hence independent of  player's beliefs).

\begin{proposition}\label{prop:PD} For the Prisoner's Dilemma of the form described
in Section~\ref{sec:dilemma}, cooperation is consistent for a player of
type
$t$ and mixed strategy $\sigma$ for the other player iff
$t \ge c$.
\end{proposition}

\prf Switching from cooperating to defecting gives the player an
additional payoff of $c$, independent of whether the other player is
cooperating or defecting.  Thus, cooperation is consistent if $t \ge
c$.
\eprf

We next consider the Traveler's Dilemma.

\begin{proposition}\label{prop:TD} For the Traveler's Dilemma,
\begin{itemize}
\item[(a)] cooperation is
  consistent with $t$ and a mixed strategy $\sigma$ for the other
  player if $t \ge 2b-1$;
\item[(b)]
      there exists a strategy $\sigma$ for
  the other player such that cooperation is
  consistent with $t$ and $\sigma$ 
iff $t \ge 2b-1$;
\item[(c)] cooperation is
  consistent with $(t,\beta)$ iff $t \ge \max(\beta(b-1),
b -   \beta(H-L))$.
\end{itemize}
\end{proposition}
\prf If player 2 plays $m$, then player 1's
best response is to play $m-1$ (or $L$ if $m=L$).  If $m < H$,
then player 1 gets a payoff of $m-b$ if he cooperates (i.e., plays
$H$), and could have gotten $m-1+b$ by 
making a best response (or $L$, in case $m=L$).  Thus, he can gain at
most $2b-1$ by playing a best response.  This proves part (a).  If
$m=H-1$ and $H-1 > L$, then cooperation is consistent iff $t \ge
2b-1$; this proves part (b).  Finally, if player 1 has type
$(t,\beta)$, then he believes that 2 plays $H$ with
probability $\beta$ and $L$ with probability $1-\beta$. Thus, player
1 believes his expected payoff from playing $H$ is $\beta H +
(1-\beta)(L-b)$.
The best response for player
1 is to play one of
$H-1$ or $L$. His payoff from playing $H-1$ is $\beta (H + b-1) +
(1-\beta)(L-b)$; his 
payoff from playing $L$ is $\beta(L+b) +(1-\beta)L$.
Thus, cooperation is
consistent for a player 1 of type $(t,\beta)$ iff $t \ge
\max(\beta(b-1), b-\beta(H-L))$.
\eprf

\fullv{For the Public Goods game, consistency is again independent of the
other players' strategies.}
\shortv{The next two results are proved in the full paper.}

\begin{proposition}\label{prop:PG} For the Public Goods game,
cooperation is consistent for a player $i$ of tolerance $t$ and
mixed strategy $\sigma_{-i}$ for the other players iff $t \ge (1-\rho)$.
\end{proposition}
\fullv{
\prf It is easy to see that, no matter what the other players do,
defection (contributing 0) is the best response in this game, and a
player gets a payoff that is $1-\rho$ higher if he defects than if he
cooperates.  Thus, cooperation is consistent iff $t \ge (1-\rho)$.
\eprf
}

\begin{proposition}\label{prop:BC} For Bertrand Competition with $n$ players,
cooperation for player $i$ is
  consistent with $(t,\beta)$ iff $t \ge
  \max(\beta^{n-1} (H-1), f(n)L) - \beta^{n-1}H/n$, where $f(n) =
  \sum_{k=0}^{n-1} \beta^k(1-\beta)^{n-1-k} \binom{n-1}{k}/(n-k)$.
\end{proposition}

\fullv{\prf Consider a player of type $(t,\beta)$.
If player $i$ cooperates, he will get $H/n$ if all the other players
cooperate, which happens with probability $\beta^{n-1}$; otherwise, he
gets 0.  Thus, his expected payoff from cooperation is $\beta^{n-1} H/n$.
His best response, given
his beliefs, is to 
play one of $H-1$ or $L$.  If he plays $H-1$, then his payoff is $H-1$
if all the remaining players play $H$, which happens with probability
$\beta^{n-1}$; otherwise his payoff is 0.  Thus, his expected payoff
is $\beta^{n-1} (H-1)$.
If he plays $L$, then his payoff
if $k$ players play $H$ and $n-1-k$ play $L$ is $L/(n-k)$; this
event occurs with probability $\beta^{k}(1-\beta)^{n-1-k}\binom{n-1}{k}$.
Thus, his expected payoff is $f(n) L$.  It follows that cooperation is
consistent with $(t,\beta)$ if  $t \ge  \max(\beta^n (H-1), f(n)L) -
\beta^{n-1}H/n$.
\eprf}

From here it is but three short steps to our desired result:
First, observe that, up to now, we have looked at games in isolation.
But now we want to compare tolerances in different games, with
different settings of the relevant parameters.  Intuitively, having a
tolerance of 2 in Traveler's Dilemma where $L=2$ and $H=100$ should
have a different meaning than it does in a version of Traveler's
Dilemma where payoffs are multiplied by a factor of 10, so that $L=20$
and $H=1000$.  Thus, when considering a family of related games,
rather than considering absolute tolerances, it seems more appropriate to
consider \emph{relative tolerance}.  There are many ways of defining a
notion of relative tolerance.  For our
purposes, we take a player's relative tolerance to be an
element of $[0,1]$; player $i$'s actual tolerance in a game $\Gamma$ is his
relative tolerance multiplied by the
payoff that player $i$ gets if everyone cooperates in $\Gamma$.  For
example, since the payoff obtained by $i$ if everyone cooperates 
in Traveler's Dilemma is $H$, then the actual tolerance of a player of
type $(\tilde{t},\beta)$ is $\tilde{t}H$.  (Here and elsewhere, if we
wish to emphasize that we are considering relative tolerance, we write
$\tilde{t}$, reserving $t$ for actual tolerance.)
There are other ways we could define relative tolerance.  For example, we
could multiply by the difference between the payoff obtained if everyone
cooperates and the payoff obtained if everyone defects,
or multiply by the maximum possible
social welfare.  The exact choice does not affect our results.

Second, recall that the fact that
cooperation is consistent with a given type does not mean that a
player of that type will actually cooperate.  We add an extra
component to the type of a player to indicate whether
the player will cooperate if it is consistent to do so, given his
beliefs.  We thus consider \emph{relative types} of the
form $(\tilde{t},\beta,C)$ and $(\tilde{t},\beta,D)$; such a type will
cooperate in
Traveler's Dilemma if $\tilde{t}H \ge \max(\beta(b-1), b -
\beta(H-L))$
and the third component is $C$.
Finally, we need to assume that there are a reasonable number of
players of each type.    Formally, we assume that the set of types
of each player is infinite and that there is a
distribution on relative types such that for all intervals $(u,v)$ and
$(u',v')$ in $[0,1]$, there is a positive probability of finding
someone of relative type $(\tilde{t},\beta,C)$ with $\tilde{t} \in
(u,v)$ and $\beta 
\in (u',v')$.  An analogous assumption is made by Capraro and
Halpern \citeyear{CH2014}.

With these assumptions, it follows from
Propositions~\ref{prop:PD}--\ref{prop:BC} that
the regularities discussed in Section~\ref{sec:dilemma} hold.
\begin{itemize}
\item In the case of Prisoner's Dilemma, $b-c$ is the payoff obtained
  if everyone cooperates,
  so if $\tilde{t}$ is the relative tolerance, $\tilde{t}(b-c)$ is the
  actual tolerance.  Thus, if a player's relative type is $(\tilde{t},\beta)$,
  then cooperation is consistent if $\tilde{t}(b-c) \ge c$.
  Clearly, as $b$ increases, there are strictly more relative types for which
  cooperation is consistent, so, by our assumptions, we should see
  more cooperation.    Similarly, if $c$ increases (keeping $b$
  fixed), there are fewer relative types for which cooperation is
  consistent, so we should see less cooperation.
\item In the case of Traveler's Dilemma, as we have observed a
  relative type will cooperate if $\tilde{t}H \ge \max(\beta(b-1), b -
\beta(H-L))$.  Clearly, if $b$ increases, then there will be fewer
relative types for whom cooperation is consistent.
\item In the Public Goods game, if everyone cooperates, the payoff to
  player $i$ is $n\rho$.  So it is consistent to cooperate if $\tilde{t}\rho n
  \ge (1-\rho)$.  Clearly, as $n$ increases, we should see more
  cooperation, given our assumptions.
Moreover, tolerance explains the increase of cooperation as the
marginal return increases.
\item Finally, in the Bertrand Competition, since the payoff if
everyone cooperates is $H/n$, it is consistent to cooperate if $\tilde{t}H/n
  \ge   \max(\beta^{n-1} (H-1), f(n)L) - \beta^{n-1}H/n$, or
  equivalently, if
$$\tilde{t} \ge \max(n\beta^{n-1} (H-1)/H, nf(n)L/H) - \beta^{n-1}.$$
  Clearly,   cooperation decreases if $L$ increases.  The effect of
    increasing $n$ is more nuanced.  For $n$ large,
$\beta^{n-1}$ is essentially 0, as is $n\beta^{n-1}$; it can be shown
    that $f(n)$ is roughly $1/(1-\beta) n$.  Thus, if $n$ is
  large, cooperation is consistent if $\tilde{t} > L/(1-\beta)H$.
  What happens for small values of $n$ is
  very much dependent on $\beta$, $H$, and $L$.  The actual
  experiments on this topic \cite{DG00} considered only  2, 3, and 4
  players, with $L=2$
  and $H = 100$.  For these values, we get the desired effect if
  $\beta$ is sufficiently large ($\beta > .7$ suffices).
\end{itemize}

As we said earlier,
of all the approaches to explaining social dilemmas in the literature,
only Capraro and Halpern \citeyear{CH2014}
and Charness and
Rabin~\citeyear{Ch-Ra}, can explain all these regularities;
see Capraro and Halpern \citeyear{CH2014} for a
detailed discussion and a comparison to other approaches.
Of course, this leaves open the question of which approach is a better
description of what people are doing.  We suspect that translucency,
care for others, and tolerance all influence behavior.
We hope
that further investigation of social 
dilemmas will reveal other regularities that can be used to compare
our approach to others,
and give us a more fine-grained understanding of what is going on.
\section{Prisoner's Dilemma with Tolerance}\label{sec:PD}

We now take a closer look at the impact of tolerance  on perhaps the
best-studied social dilemma, Prisoner's Dilemma.
The analysis suggests how thinking in terms of tolerance might help us
design better mechanisms.

The general prisoners' dilemma (PD) game has payoffs $(a_1,a_2)$,
$(b_1,c_2)$, $(c_1,b_2)$, and $(d_1,d_2)$ corresponding to
action profiles $(C,C)$, $(C,D)$, $(D,C)$, and $(D,D)$, respectively,
with $c_i  > a_i > d_i > b_i$. 
%
To analyze equilibrium outcomes in PD with tolerances,
consider a
player $i$, and suppose she believes that the other player $j$ will
play $C$ with probability $\alpha_j$. Her payoff from choosing action $C$
and action $D$ are, respectively,
\BEQ
\nonumber
u_C = a_j \alpha_i + (1-\alpha_j) b_i; ~~ u_D = \alpha_j c_i +
(1-\alpha_j) d_i.
\EEQ
Since $D$ is a dominant strategy in Prisoner's
Dilemma, $u_D > u_C$.  
Agent $i$ is willing to play $C$ if $u_D-u_C$ 
is within her tolerance, that is, if 
\BEQ
\nonumber
t \geq \alpha_j (c_i - a_i) + (1-\alpha_j) (d_i - b_i) \triangleq
\alpha_j \Delta C_i + (1-\alpha_j) \Delta D_i, 
\EEQ
where $\Delta C_i$  is the gain to player $i$ from defecting
when the the other player plays C, and similarly for $\Delta
D_i$.
Taking $F_i$ to denote the cumulative probability distribution on
agent $i$'s tolerances, it follows that
the probability that agent $i$ has the tolerance required to
allow cooperation is $1-F_i(\alpha_j \Delta C_i + (1-\alpha_j) \Delta
D_i)$.

Note that the minimum tolerance at which $i$ can cooperate, which
depends upon the probability $\alpha_j$ with which $j$ plays C, need
{\em not} decrease with $\alpha_j$: If the payoffs in the game are
such that $\Delta C_i > \Delta D_i$ (the gain from defecting is larger
when the other player is cooperating rather than defecting), then
increasing $\alpha_j$ {\em increases} the tolerance $i$ must have to
be willing to cooperate.

Suppose that agents break ties in favor of cooperation, that is, if
cooperating yields a payoff within an agent's tolerance, that agent will
cooperate rather than defect.
Call a $\pi$-tolerant equilibrium satisfying this condition a
\emph{particularly cooperative} ($\pi$-tolerant) equilibrium.  
Note that if some player with tolerance $t$
cooperates in a perfectly cooperative equilibrium, then all players
with tolerance $t$ cooperate.

It follows from the discussion above that a particularly cooperative
equilibrium is determined by a pair 
of mutually consistent probabilities of cooperation $(\alpha_i,
\alpha_j)$ satisfying 
$$\begin{array}{l}
  \alpha _i = 1-F_i(\alpha_i \Delta C_j + (1-\alpha_i) \Delta D_j)\\
  \alpha _j = 1-F_j(\alpha_j \Delta C_i + (1-\alpha_j) \Delta D_i).
\end{array}
$$

Note that a particularly 
cooperative equilibrium may not exist in PD, 
although a $\pi$-tolerant equilibrium always does (since both agents
defecting is a $\pi$-tolerant equilibrium, no matter what $\pi$ is).
For a simple example,
suppose that $a_1 = a_2 = 3$,  $b_1 = b_2 = -1$, $c_1 = c_2 = 5$,
$d_1 = d_2 =0$, and players are drawn from a population where everyone
has a tolerance of 1.5.
Now suppose
that there is a perfectly cooperative equilibrium where a fraction
$\alpha$ of the players cooperate.  Thus, a player's
expected payoff
from cooperation is $3\alpha -(1-\alpha) = 4\alpha - 1$; a player's
expected payoff from defection is $5\alpha$.  Thus, a player gains
$\alpha + 1$ by defecting.  If $0 \le \alpha \le .5$, then a player
gains
at most $1.5$ by switching from cooperate to defect, so all
players should cooperate in a perfectly cooperative equilibrium (i.e.,
$\alpha$ should be 1); on
the other hand, if $\alpha > .5$, then $1 + \alpha > 1.5$, so all
players should defect (so $\alpha$ should be 0).

In this example, we have a point mass of 1 on  
tolerance 1.5, so there is only one type of each player.  This is
inconsistent with the assumption in the previous section that the
cumulative probability increases continuously.  If we assume that
the cumulative probability increases continuously then there is always a
particularly cooperative equilibrium in PD.
We provide an analysis here, making some symmetry
assumptions for simplicity. 
Specifically, we assume
 (1) symmetry in payoffs: $a_1 = a_2, \ldots, d_1 = d_2$; 
(2) symmetry in tolerance distributions: $F_1 = F_2 = F$; and
(3) that $F$ is continuous (so that there are infinitely many types).
Under these assumptions, we show that a \emph{symmetric} perfectly cooperative
equilibrium (where $\alpha_1 = \alpha_2$) always exists; this is a 
solution $\alpha^*$ to
\begin{equation}\label{e-PDeq}
1- \alpha = F(\alpha \Delta C + (1-\alpha) \Delta D).
\end{equation}
(Note that for prisoners' dilemma, $0 < F(\Delta C),
F(\Delta D) \leq 1$ since $\Delta C = c-a$ and $\Delta D = d-b$ are
positive.)

\begin{figure}[htb]
\vspace{-.5in}
\begin{minipage}[c]{0.46\textwidth}
\begin{center}
\includegraphics[scale = .14]{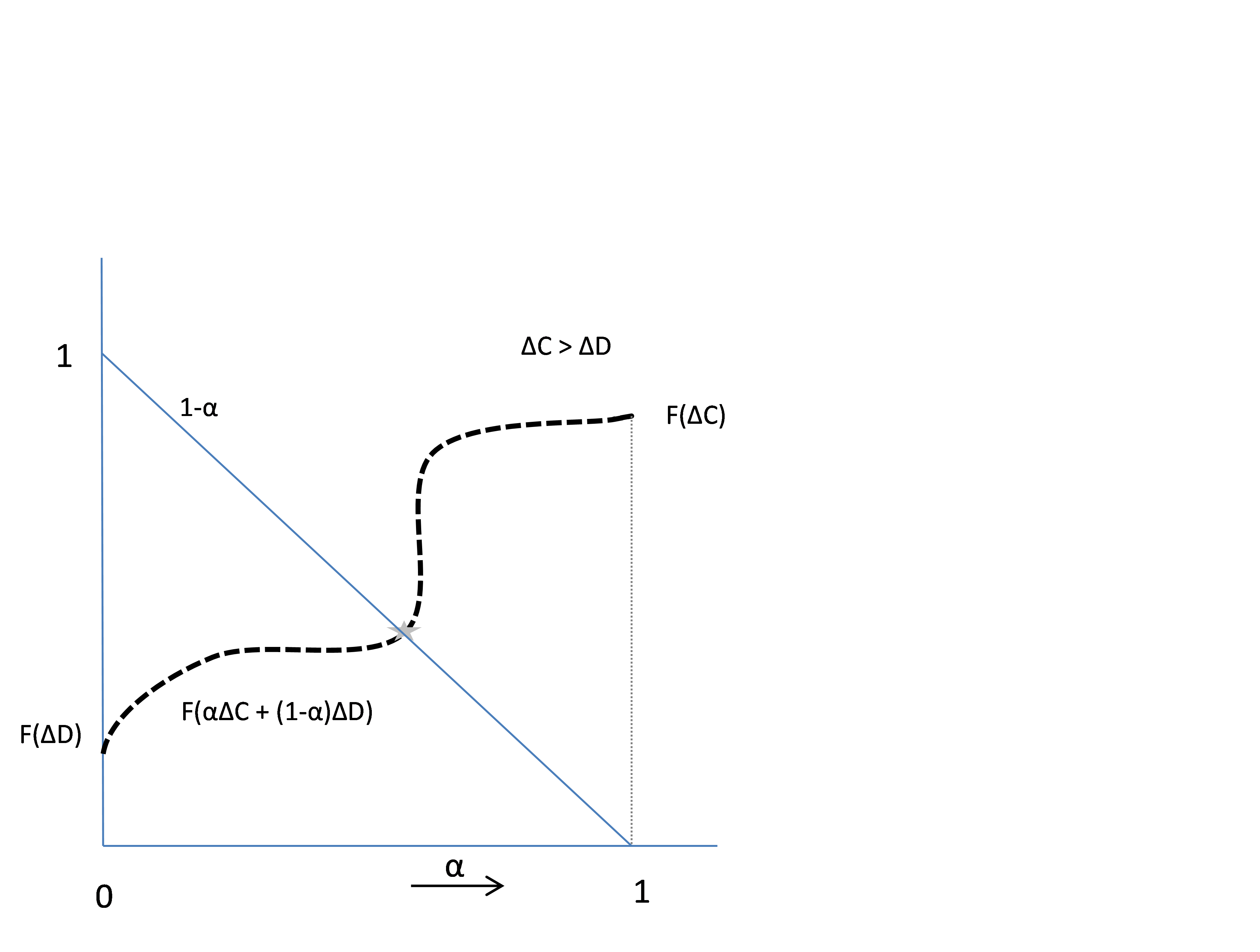}\\
(a) $\Delta C > \Delta D$
\end{center}
\end{minipage} \ \ \ \ 
\begin{minipage}[c]{0.46\textwidth}
\begin{center}
\includegraphics[scale = .14]{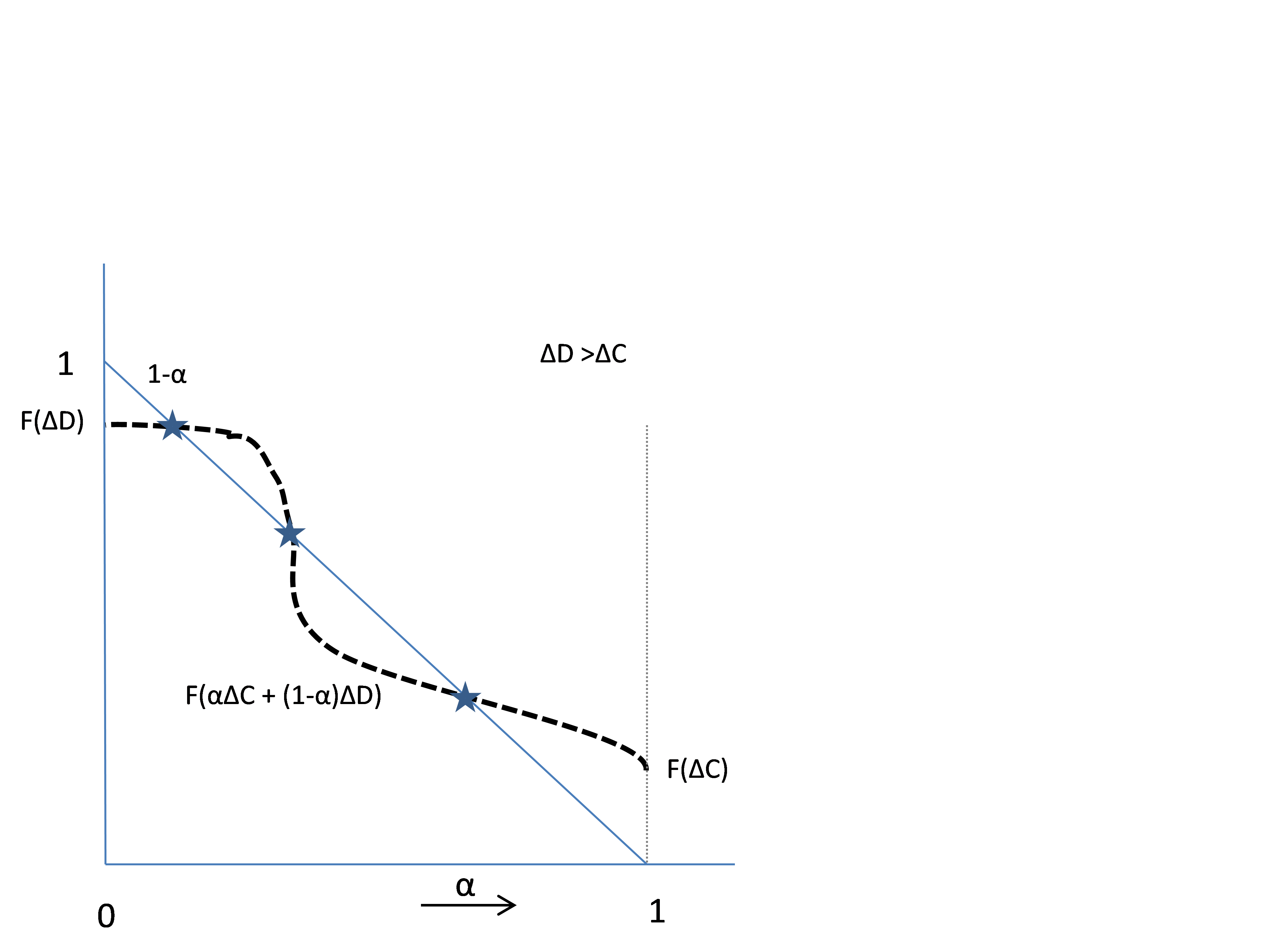}\\
(b) $\Delta C \le \Delta D$
\end{center}
\end{minipage}
\label{eqbmPD}
\caption{Equilibrium structure for PD.}\label{f-eqbmPD}
\end{figure}

\begin{theorem}[Equilibrium structure.] Under our assumptions, a
  symmetric particularly cooperative
  equilibrium
  always exists,
  is a solution to (\ref{e-PDeq}),
  and has the following structure:
\BNUM
\item[(a)] There is an equilibrium with $\alpha = 0$ (in which case $(D,D)$ is
  necessarily played, so there is no cooperation) 
 if and only if $F(\Delta D) =1$.
\item[(b)] (Uniqueness.) If $\Delta C > \Delta D$, there is a unique
equilibrium; if $\Delta C \leq\Delta D$, multiple equilibria
  corresponding to different cooperation probabilities may exist.
\ENUM
\end{theorem}
We omit the formal proofs; the results follow from applying the
Intermediate Value Theorem using the continuity of $F$, and noting that the
LHS in (\ref{e-PDeq}) is greater than the RHS at $\alpha = 0$ when
$F(\Delta D) < 1$, and is smaller than the RHS at $\alpha = 1$.
Uniqueness (and non-uniqueness, respectively) follows from the fact
that the RHS is increasing (respectively, decreasing) in $\alpha$ when
$\Delta D < \Delta C$ (respectively $\Delta C < \Delta D$). The idea
is perhaps best illustrated by Figure \ref{f-eqbmPD}, where the
intersections correspond to equilibrium cooperation probabilities
$\alpha^*$.
Note that these results depend critically on $F$, the cumulative
distribution, being continuous.  


The next result gives insight into
how the probability of cooperation changes as we change various
parameters.  As  we change the relevant
parameters  (the payoffs and the probabilities of tolerance)
slightly in a continuous way,
each particularly cooperative equilibrium ``shifts'' slightly also in
a continuous way, so that we can talk about corresponding equilibria;
we omit the formal definition here.
\begin{theorem}\label{t-cs}
The equilibrium probability of cooperation $\alpha^*$
in corresponding particularly cooperative equilibria
(a) decreases as $\Delta C$ and $\Delta D$ increase, and therefore (b) decreases 
as the payoffs $c$ and $d$ increase, and increases as $a$ and $b$ increase,
and (c) ``increases'' in $\pi$, in that if $\pi'$ stochastically
dominates $\pi$, then the payoffs in a particularly cooperative
$\pi'$-tolerant equilibrium are higher than in the corresponding $\pi$-tolerant
equilibrium.
\end{theorem}
These results follow easily by noting that
for a fixed $\alpha$, the RHS  in (\ref{e-PDeq})
is increasing in both $\Delta C$ and $\Delta D$; the value of $\alpha$
at which the RHS equals the LHS therefore decreases when either of
$\Delta C$ and $\Delta D$ increase. The monotonicity in $\pi$ is
similar: if $\pi'_1$ stochastically dominates
$\pi_1$, the
value of $\alpha$ at the intersection is larger with  $\pi'_1$
than with $\pi_1$.

These results, in addition to providing testable predictions for how
cooperation levels should behave when payoffs are varied in an
experiment, also provide guidelines for a designer who may be able to
manipulate some of the payoffs in the game (via either social or
monetary means), by isolating
the factors that influence the nature of equilibria and extent of
equilibrium cooperation. First, the extent of cooperation in
equilibrium depends on the marginal, rather than the absolute,
payoffs: it is the {\em differences} $\Delta C$ and $\Delta D$ that
determine equilibrium levels of cooperation, rather than any other
function of the payoffs $a,b,c,d$.%
\footnote{We must be a little careful here.  To do
  comparative statics, we should consider relative tolerances, for the
  reasons discussed in Section~\ref{sec:dilemma}.  Changing the payoff
  parameters may well change the actual tolerance, while keeping the relative
  tolerance fixed.  Parts (a) and (b) of Theorem~\ref{t-cs} hold only for
  a fixed tolerance distribution.} 

Second, perhaps surprisingly,
which of $\Delta C$ or $\Delta D$ is larger makes a difference to
the structure of equilibria, which also has implications for
design. If, for instance, the designer prefers a game where there is a
unique equilibrium with non-trivial cooperation, Theorem \ref{t-cs}
suggests that the designer should manipulate payoffs so that $\Delta
D$, the marginal gain from defecting instead of cooperating when the
other player also defects, is smaller than  $\Delta C$, the marginal
gain from defecting when the other player cooperates. (This might be
achieved, for instance, by providing additional rewards---either extra
compensation, or social rewards such as public acknowledgement, to a
player who continues to cooperate despite defection by her partner,
increasing the payoff $b$ and therefore decreasing
$d-b$.) On the other hand, if there is a means to ``nudge'' behavior
towards a particular equilibrium when there are multiple equilibria, a
designer might prefer to manipulate payoffs to fall in the $\Delta C
\leq \Delta D$ regime and nudge behavior towards the equilibrium with 
the most cooperation (\fullv{again, this could be achieved }by imposing
social or monetary penalties for defecting on a cooperating partner,
decreasing $t$ and thereby $\Delta C$).


\section{Conclusion}
\label{sec:conclusion}

We have defined a notion of $\pi$-tolerant equilibrium, which takes
into account that players have some tolerance
regarding payoffs.  This solution concept generalizes Nash
and $\epsilon$-Nash equilibrium in a natural way.  We showed that this
solution concept can explain cooperation in social dilemmas.
Although we focused on social dilemmas, tolerance can also explain
other well-known observations, such as the fact that people give some
money to the other person in
the \emph{Dictator Game} \cite{KKT86} (where one person is given a
certain amount of money, and can split it as he chooses between
himself and someone else) and that people give intermediate amounts
and reject small positive offers in the
\emph{Ultimatum Game} \cite{GSS82} (where one person can decide on how
to split a certain amount of money, but the other person can reject 
the split, in which case both players get nothing).

We also examined
the structure of particularly cooperative $\pi$-tolerant equilibria,
where players are as cooperative as they can be,
given
their tolerances, in Prisoner's Dilemma. To the extent that cooperation is
due to tolerance, our results provide guidance to a mechanism designer
who has some control over the payoffs in a game, and suggest ways that
cooperation can be increased.  Since many practical situations of
interest can be modeled as Prisoner's Dilemmas, these results may
suggest how mechanism designers can take advantage of players'
tolerance in practice.

We believe that a study of convergence towards, and stability and
robustness of, particularly cooperative equilibria in Prisoner's
Dilemma in an appropriate  model for dynamics can potentially provide
useful insights into emergence and sustainability of trust in online
economies.

\appendix

%
\section{Proof of Theorem 2.3}

\newenvironment{RETHM}[2]{\trivlist \item[\hskip 10pt\hskip\labelsep{\sc #1\hskip 5pt\relax\ref{#2}.}]\it}{\endtrivlist}
\newcommand{\rethm}[1]{\begin{RETHM}{Theorem}{#1}}
\newcommand{\erethm}{\end{RETHM}}
In this section, we prove Theorem~\ref{thm:stochastic}.  We repeat the
statement of the theorem for the reader's convenience

\rethm{thm:stochastic} If $\pi'$ stochastically
  dominates $\pi$, then 
every $\pi$-tolerant equilibrium is a $\pi'$-tolerant equilibrium.
\erethm

Suppose that $\sigma$ is a $\pi$-tolerant equilibrium, and
$\pi'$ stochastically dominates $\pi$ for all $i$.  We want to show that $\sigma$
is a $\pi'$-tolerant equilibrium.
Clearly, it suffices to consider the case where $\pi_i'$ dominates
$\pi_i$, and $\pi_j' = \pi_j$ for $j \ne i$.

Let the support of $\pi_i$ be $\{t_1, \ldots, t_n\}$, where $t_1 <
\ldots < t_n$, and let the support of $\pi_1'$ be $\{t_1', \ldots,
t_m'\}$, where $t_1' < \ldots < t_m'$.  For
convenience, define $t_0 = t_0' = 0$.
By assumption, there exists a mapping $g_i$ such that $g_i(t)$
is a mixed strategy for each type $t$ with support consisting of pure strategies
consistent with $t$ and $\sigma_{-i}$ (this is E1) such that
$\sum_{h=1}^n \pi_i(t_h) g_i(t_h) = \sigma_i$ (this is E2).  We want to
define a comparable function $g_i'$.
In the remainder of the proof, for ease of exposition, we drop the
subscript $i$ (on $g_i$, $g_i'$, $\pi_i$, $\pi_i'$).

We start by defining $g'(t_1')$. This has the benefit of
giving the intuition for how to define $g'$ in general.
For ease of notation, we write $F$ and $F'$ rather than
$F^{\pi}$ and $F^{\pi'}$.
Choose the smallest $h$
such that $F(t_h) \ge F'(t_1')$.
Intuitively, type $t_1'$ should play
$g(t_1)$ with probability $\pi(t_1)/\pi'(t_1')$,
$g(t_2)$ with probability $\pi(t_2)/\pi'(t_1')$,
\ldots,
and $g(t_h)$ with probability $\pi(t_h)/\pi'(t_1')$.
This isn't quite right, since $\sum_{j=1}^h
\pi(t_j) = F(t_h)$, and $F(t_h)$ may be strictly larger than
$F'(t_1) = \pi'(t_1)$.  Thus, we modify the probability that
$t_1'$ plays $g(t_h)$ to
$(\pi'(t_1) - F(t_{h-1}))/\pi'(t_h')$.
That is, $g(t_h)$ is played with whatever probability is left over after
all the other strategies have been played.  
Thus, we take $g_1'(t_1')$, the strategy played by type $t_1'$, to be 
\begin{equation}\label{eq1}
\frac{\pi(t_1)}{\pi'(t_1')} g(t_1) + \cdots +
\frac{\pi(t_{h-1})}{\pi'(t_1')} g(t_{h-1}) +
\frac{\pi'(t_1) - F(t_{h-1})}{\pi'(t_1')} g(t_h).
\end{equation}

We must show that E1 is satisfied by $\pi'$,
so that any pure strategy in the
support of any of $g(t_1), \ldots, g(t_h)$ is consistent with $t_1'$
and $\sigma_{-i}$.
Since consistency is monotonic in the tolerance, and by E1, all the
strategies in the support of $g(t_j)$ are consistent
with $t_j$ and $\sigma_{-i}$, it suffices to
show that $t_1' \ge t_h$.  Suppose, by way of contradiction, that $t_1' < t_h$.
By choice of $t_h$, if $t < t_h$, then $F(t) < F'(t_1')$.  Thus,
$F(t_1') < F'(t_1')$.  But this contradicts the assumption that
$F'$ stochastically dominates $F$.
Thus, E1 is indeed satisfied by $\pi'$.


We now define $g'(t_j')$ for $j \ge 2$.
We first define auxiliary functions $\alpha:\{1,\ldots,m\} \rightarrow
\{1,\ldots,n\}$ and $\beta:\{1,\ldots,m\} \rightarrow [0,1]$:
\begin{itemize}
\item $\alpha(j)$ is the least $h$ such that $F(t_h) \ge
F'(t_j')$ (so the $h$ in the definition of $g'(t_1')$ above is just
$\alpha(1)$).
\item $\beta(j)$ is defined by induction on $j$.  Let $\beta(1) =
\pi'(t_1) - F(t_{\alpha(1)-1})$.  
Note that $\beta(1)$ was the quantity
that occurred in the argument above that represented 
the amount of the probability mass $\pi'(t_1')$ not allocated to $g(t_1), \ldots,
g(t_{\alpha(1)-1})$, which can thus be allocated to $g(t_{\alpha(1)})$.
For $j \ge 2$, define
$$\beta(j) = \pi'(t_j) - \left(\pi(t_{\alpha(j-1)}) - \beta(j-1) +
\sum_{h=\alpha(j-1)+1}^{\alpha(j)-1}\pi(t_h)\right). $$
Again, roughly speaking, $\beta(j)$ is the amount of the probability mass
$\pi'(t_j')$ not allocated to the strategies $g(t_{\alpha(j-1)}), \ldots,
g(t_{\alpha(j)})$.
\end{itemize}
We claim that, for $j \ge 2$, $t_{j}' \ge t_{\alpha(j)}$.  The argument is
essentially the
same as that given above for the case that $j=1$.  Suppose not.  Then, by
choice of $t_{\alpha(j)}$, if $t < t_{\alpha(j)}$, then $F(t) < F'(t_{j}')$.  Thus,
$F(t_{j}') < F'(t_{j}')$.  But this contradicts the assumption that
$F'$ stochastically dominates $F$.

Finally, for $j > 1$, define
\begin{equation}\label{eq2}
g'(t_j') =
\frac{1}{\pi_1'(t_j')}\left[ (\pi(t_{\alpha(j-1)}) -\beta(j-1)) g(t_{\alpha(j-1)}) +
\left(\sum_{h =  \alpha(j-1)+1}^{\alpha(j) -1} \pi(t_h) g(t_h)\right) +
\beta(j)g(t_{\alpha(j)}) \right].
\end{equation}

To see that this works, we need to check E1 and E2.  For E1, note that
the support of $g'(t_j')$ consists of the strategies in the support of 
$g(t_{\alpha(j-1)}), \ldots, g(t_{\alpha(j)})$.  Since strategies in the support of
$g(t_h)$ are all consistent with $t_h$ and $\sigma_{-i}$, all the
strategies in the support of $g'(t'_j)$ are clearly consistent with
$t_{\alpha(j)}$ and $\sigma_{-i}$.  We showed above that $t_{\alpha(j)} \le
t_j'$, so all these strategies are consistent with $t_j'$ and
$\sigma_{-i}$.  Thus E1 holds.

For E2, note that it is immediate from (\ref{eq1}) and the definition
of $\alpha(1)$ and $\beta(1)$ that
\begin{equation}\label{eq3}
\pi_1'(t_1') g_1'(t_1') = \sum_{h=1}^{\alpha(1)-1} \pi(t_h) g(t_h) +
\beta(1)g(t_{\alpha(1)}).\end{equation}
It is immediate from (\ref{eq2}) that for $j > 1$, 
\begin{equation}\label{eq4}
\pi_1'(t_j') g_1'(t_j') = 
(\pi(t_{\alpha(j-1)}) -\beta(j-1)) g(t_{\alpha(j-1)}) +
\left(\sum_{h =  \alpha(j-1)+1}^{\alpha(j) -1} \pi(t_h) g(t_h)\right) +
\beta(j)g(t_{\alpha(j)}).\end{equation}  It follows from (\ref{eq3}) and
(\ref{eq4}) that 
$$\sum_{h=1}^m \pi'(t_h') g'(t_h') =
\sum_{h=1}^n \pi(t_h) g(t_h).$$  Since the latter sum is $\sigma_1$
by E2, we are done.
\eprf

  \paragraph{Acknowledgments:} We thank Valerio Capraro for useful
comments on an earlier draft of the paper.  The work of Arpita Ghosh
was supported in part by NSF grant CCF-1512964 and ONR grant
N00014-15-1-2335 the work of Joe Halpern was supported in 
part by NSF grant CCF-1214844, AFOSR grant
FA9550-12-1-0040, and ARO grants W911NF-14-1-0017 and W911NF-16-1-0397.

\bibliographystyle{eptcs}
\bibliography{joe}
 
\end{document}